\documentclass[aps,prb,superscriptaddress,twocolumn,floatfix,showpacs,amsmath]{revtex4-1}
\usepackage{graphicx}
\graphicspath{ {./images/} }
\usepackage{subcaption}
\usepackage{floatrow}
\usepackage{latexsym}
\usepackage{psfrag}
\usepackage[makeroom]{cancel}
\usepackage{dcolumn}
\usepackage{blindtext}
\usepackage{epstopdf}
\usepackage{float}
\floatstyle{plaintop}
\restylefloat{table}

\usepackage{amsmath,latexsym,psfrag}
\usepackage{array}
\usepackage{dcolumn}
\usepackage{bm}

\usepackage{soul}  

\usepackage[%
colorlinks=true,
urlcolor=blue,
linkcolor=blue,
citecolor=blue
]{hyperref}
\begin{document}

	\title{Interplay of lattice distortion and bands near the Fermi level in $A$TiO$_3$  ($A$=Ca, Sr, Ba)}
	\author{Patrick Ning'i}
	\email{ningipatrick@gmail.com }
	\affiliation{ Materials Modeling Group, 
		School of Physics and Earth Science,
		The Technical University of Kenya,
		52428-00200, Nairobi, Kenya.}  
	\author{Stephen Chege}
	\affiliation{ Materials Modeling Group, 
		Department of Physics and Space Sciences,
		The Technical University of Kenya,		52428-00200, Nairobi, Kenya.}  
	
	\author{ James Sifuna}
	\affiliation{ Materials Modeling Group, 
		Department of Physics and Space Sciences,
		The Technical University of Kenya,
		52428-00200, Nairobi, Kenya.}  
	\affiliation{ Department of Natural Science, 
		The Catholic University of Eastern Africa,
		62157 - 00200, Nairobi, Kenya.}
	\author{George Amolo}
	\affiliation{ Materials Modeling Group, 
		Department of Physics and Space Sciences,
		The Technical University of Kenya,
		52428-00200, Nairobi, Kenya.} 
	
	\date{\today}
	
	\begin{abstract}
    The structural and electronic properties of $A$TiO$_3$ ($A$=Ca, Sr, Ba) have been investigated under strain-free situation and with realistic constraints using first-principles calculations. We endeavored to find out the interplay between mild lattice distortions and bandgap in three $A$TiO$_3$ family members that has remained skeptical to date. We found out that the electronic structure was particularly sensitive to strains (compressive or tensile) as expected in most materials science studies. Our results indicate that under mild strains; the bandgap ($E_{gap}$), increased under compression and decreased under tension. In all the three materials, the bandgap and the lattice parameter ($a$) were found to relate as $E_{gap}\propto \frac{1}{a^x}$ for mild distortions with $2.19<x<3.1$. All these changes are attributed to the interplay of electrostatics and covalency in these crystals. This work acts as a yardstick on bandgap engineering to achieve desired properties in these titanates for feasible future  applications.

	\end{abstract}
	

	\maketitle
	\section{INTRODUCTION}
	\label{sec:introduction}
	Transition metal oxides (TMOs) have catapulted interest in both academia and industry for the last over fifty years. The ever growing interest is similar to that of semiconductor physics that came in to play many years ago~\cite{oxides_meet,Aguado-thesis}. We believe that this comparison is valid in the sense that our lives today are highly inclined on devices developed from fundamental materials science research.
	It is important to note that most transition metals oxides are coupled by strong correlations that give birth to novel physics; ferromagnetism, ferroelectricity and superconductivity among others. Novel physics in TMOs arise from a perfect interplay of charge, lattice, orbital and spin degrees of freedoms~\cite{physics_of_oxides,Tokura462,1DMFT,2oxides,ghosez2006firstprinciples}.

	Perovskites are a family of materials in the TMOs with a chemical formula ABO$_3$. With five atoms in the unit cell in their high-symmetry cubic phase, they contain new and magnificent physics. Materials in this class all have similar atomic structures~\cite{Aguado-thesis}.
	The structural stability in the ABO$_3$ class is remarkable for a wide choice of $A$ and $B$ cations~\cite{Catalano_2018_long} and is determined by the famous Goldschmidt tolerance factor (t)~\cite{Goldschmidt}. In our case, we employed perovskites containing a transition metal on the B-site and thus anticipate to have numerous electronic phases arising. The novelty arises from well known complex physics of $d$ electrons and the fact that their states overlap or at times they overlap with the  O-2$p$ states. All these can be tuned by minimal structural changes which are usually seen as perturbations.

	Here we give a focus on \textit{A}TiO$_3$ perovskites. These materials have intrinsic properties that are tied on their atomic constituent. It has been found out that perovskite-like oxides in which CaTiO$_3$, SrTiO$_3$ and BaTiO$_3$ are embedded, work efficiently as catalysts due to their high melting points and also provide successful oxygen flow carriers~\cite{importance1}. Elsewhere, the non-centrosymmetric structures of these materials may possess ferroelectric properties~\cite{importance_2,importance_4,importance_3} and thus implies great application in electro-optic switching devices. 
	As seen in the papers referenced herein, a number of first principles calculations for the ABO$_3$ compounds have been reported in recent years. However, some questions regarding the electronic properties and their dependency on lattice distortions still linger. 

   Many years ago, Dalven~\cite{delvin} deduced an empirical relation $E_0\propto \frac{1}{a_0^2}$, between the energy gap $E_0$ and lattice constant $a_0$ for a wide range of semiconductors with the cubic NaCl crystal structure. These findings dictated that applied strain on a crystal always affected its intrinsic properties~\cite{scf3} and thus leading to subtle physics. This motivated us to test Dalven's findings on three members in the $A$TiO$_3$ family to see if a similar trend can be realised on strain application. 

     Many scholars~\cite{4,5,6,7,8,Yang_Wang_Chen_Yan_2016} have reported strain-induced structural and electronic property modulations of ZnA (A=O, S, Se and Te), GdN and AN (A=Al, Ga) employing first principles calculations. Bousquet and co-workers~\cite{8}, demonstrated theoretically that by application of appropriate strains, ferroelectricity can be induced in alkaline-earth metal binary oxides like BaO. 
     
    %
    
    In this work, the authors intend to investigate the effects of strains on the structural and electronic properties of \textit{A}TiO$_3$ by performing first principles calculations. The output from this work may be of help in revealing the driving mechanism in the relationship between lattice and bandgap as reported by Dalven~\cite{delvin} , thus paving way for new applications. 

    The remaining parts of the paper are organized as follows:
    In Sec.~\ref{sec:technicalities} we account for the technicalities in our calculations.
    In Sec.~\ref{sec:results}, we present the structural properties of the three titanates in comparison to other works (Sec.~\ref{sec:structure}), the electronic structure in an unstrained case (Sec.~\ref{sec:equilibrum}), the various active orbital contributions (Sec.~\ref{sec:orbital}) and the response of the bandgap to various lattice strains (Sec.~\ref{sec:lattice_distortion}).
    Conclusion and future perspectives regarding the work are given in Sec.~\ref{sec:conclusions}.	
	\section{TECHNICALITIES}
	\label{sec:technicalities}
	The simulations were done on \textit{A}TiO$_3$, (A=Ca, Sr, Ba) by employing the numerical atomic orbital method as implemented in the 
	{\sc Siesta} method~\cite{SIESTA}.
	Exchange  and correlation functions were treated by the generalized gradient approximation (GGA)~\cite{PBE-1996} to the density functional theory (DFT)~\cite{Hohenberg-64,Kohn-65}. 
	It is important to mention that only intrinsic properties of the titanates were considered in these calculations and thus no  Hubbard-term to deal with the on-site Coulomb repulsion on the Ti $d$ states was considered.
	In all the calculations herein, we replaced  the core electrons by {\it ab initio} norm conserving pseudopotentials that followed the Troullier-Martins scheme~\cite{Troullier-91} in the Kleinman-Bylander fully non-local separable representation~\cite{Kleinman-82}.
	Since there is a large overlap between the semi-core states and valence states, the 3$s$ and 3$p$ electrons of Ca, 4$s$ and 4$p$ electrons of Sr, 5$s$ and 5$p$ electrons of Ba and 3$s$ and 3$p$ electrons of Ti were explicitly included in the calculations. All the pseudopotentials in this calculations  were generated in a scalar-relativistic format. The chosen configuration and cutoff radii for each shell for the pseudopotentials generated in this work can be found in Ref.~\onlinecite{Junquera-03.2} for Sr, Ba,  Ti and O,  and in the Table~\ref{table:pseudopotentials} for Ca.
	
	\begin{table}[h]
		\caption[ ]{ Chosen configuration and cutoff radii (in bohr)
			for Ca pseudopotential employed in our study.
		}
		\begin{center}
			\begin{tabular}{cccc}
				\hline
				\hline
				Reference                         &
				&
				$3s^{2}, 3p^{6}, 3d^{0}, 4f^{0}$  \\
				\hline
				Core radius                       &
				$s$                               &
				1.50                              \\
				&
				$p$                               &
				1.50                              \\
				&
				$d$                               &
				1.90                              \\
				&
				$f$                               &
				2.00                              \\
				&
				Scalar relativistic?              
				&
				yes                               \\
				\hline
				\hline
			\end{tabular}
		\end{center}
		\label{table:pseudopotentials}
	\end{table}

	In {\sc {Siesta}}, we expanded the one-electron eigenstates in a set of strictly localized numerical atomic orbitals~\cite{Sankey-89,Artacho-99}. A Fermi-Dirac smearing distribution with a temperature of 870 K was used to smear the occupancy of the one-particle electronic eigenstates
	In this calculation, We employed  a single-$\zeta$ (SZ) basis set for the semicore states of Ca, Sr, Ba and Ti, and double-$\zeta$ plus polarization (DZP) for the valence
	states of all the atoms.
	Using plane-wave cutoff of 600 Ry in the representation of charge density, we were able to calculate the corresponding matrix elements between the orbitals, the electronic charge density, the Hatree potentials together with the exchange-correlation potential. 

	In order to obtain a converged system, a two step procedure was performed: Step one, to relax the atomic structure and the one particle density matrix using a sensible number of k-points (6$\times$6$\times$6 Monkhorst-Pack~\cite{Monkhorst-76} k-point mesh) and step two, freezing the already relaxed structure and density matrix, a non-self consistent band structure calculation was performed using a much denser sampling of 60$\times$60$\times$60 in the real space integrations.  In this calculation, the atomic coordinates were  relaxed until the forces were smaller than $0.01$ eV/\AA ~and we ensured that the stress tensor components were below 0.0001 eV/\AA$^3$.
	
	\begin{figure}[H]
		\begin{center}
			\includegraphics[width=0.6\textwidth]{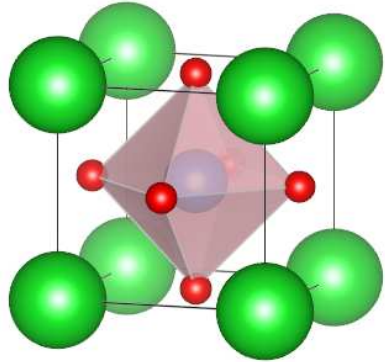}
			\caption{(Color online) The schematic representation of the unit cell of  titanates in this study, the green represents A-cation (Ca, Sr, Ba),  the blue colour inside the octahedra represents the B-cation (Titanium) while the red represents the oxygen atoms.  By a simple inspection, we can tell that the A-cation portrays a 12-fold oxygen coordination while B-cation shows a 6-fold oxygen coordination, forming the BO$_6$ octahedara which is a critical signature in the functional unit of perovskites. The positions of the atoms are as follows A(0.0,0.0,0.0), B(0.5,0.5,0.5), O$_1$(0.5,0.5,0.0),O$_2$(0.5,0.0,0.5), O$_3$(0.0,0.5,0.5).}
			\label{fig:equiii}
		\end{center}
	\end{figure}

	\section{RESULTS AND DISCUSSIONS}
	\label{sec:results}
	\subsection{Structure of the three titanates}
	\label{sec:structure}
	In Fig.~\ref{fig:equiii}, we show a typical representation of an $A$TiO$_3$ crystal structure.
	To calculate the structural properties, we obtained the energy volume relationship and fitted to the Murnaghan equation of state~\cite{Murnaghan244} so as to get the volume that yielded the minimum energy value. In Table~\ref{table:abulkcubic}, we compare both the calculated and experimental values of the three titanates. 
	It is important to note that the only perovskite among the three titanates which
	displays a cubic structure at room temperature is SrTiO$_3$~\cite{SrTiO3} while CaTiO$_3$ and BaTiO$_3$ do not~\cite{CaTiO3,BaTiO3}. In this section of the paper, we focus on the structural properties of ATiO$_3$, (A=Ca, Sr, Ba) so as to elucidate their similarities and differences.

	\begin{table}[H]
		\caption {
			Calculated (cal) and experimental (exp) lattice parameters
			($a$ in $\rm{\AA}$) for CaTiO$_{3}$, SrTiO$_{3}$, and
			BaTiO$_{3}$ in the bulk cubic structure. Numbers in parentheses are deviations from experiment.
		}
		\label{table:abulkcubic}
		\begin{center}
	
	\begin{tabular}{llc c c}

		\hline
		\hline
		System&$a_{\rm cal}$            &
	$a_{\rm exp}$  \\
		\hline
	CaTiO$_3$&	3.89 (5.4\%)                   &
	3.84~\cite{Rabe-07} \\
		
	SrTiO$_3$&	3.93 (2.7\%)                   &
		3.91~\cite{Rabe-07}\\
		
	BaTiO$_3$&	4.02 (2.0\%)                   &
		4.00~\cite{Rabe-07}  \\
		\hline
		\hline
	\end{tabular}
  \end{center}
\end{table}
From inspection, it can be noted that the calculated values  of the lattice parameter lie slightly above the experimental ones for the three titanates. This in principle should not raise an alarm since it's a well known problem, in that the GGA approximation will tend to overestimate the lattice parameters as discussed in Ref.~\onlinecite{aip_james}. In the three titanates, CaTiO$_3$ has the smallest lattice parameter. Again, this is intentional in that there is a huge relationship in the ionic radius of Ca, Sr, Ba and the lattice parameter. In the three cubic crystal structures of \textit{AB}O$_3$, the lattice parameter ($a$) and the ionic radii will tend to obey the relation in Eq.~(\ref{Eqn 1}). Where, $r_A$, $r_B$ and $r_O$ are the radii of A, B and C ions in that order\cite{Catalano_2018_long}.
\begin{align}
a=\sqrt{2}(r_{A}+r_O)=2(r_{B}+r_O).
\label{Eqn 1}
\end{align}
Another comparison between the three titanates is the measure of the respective bond-lengths. Bond-lengths may predict the hardness of any material. In principle, a shorter bond will be stronger than a longer bond. By inspection, there exists a relationship between the ionic radii and the lattice parameter ($a$) of the \textit{A}TiO$_3$, (\textit{A}=Ca, Sr, Ba). If one has a clear picture of the crystal in mind, then  the bonds \textit{A}-O and \textit{B}-O can be calculated from Table~\ref{table:abulkcubic} as $\frac{\sqrt{2}a}{2}$ and $\frac{a}{2}$, respectively. This is  yields Table~\ref{table:bonds}. The bonds arise between each cation and anion.

\begin{table}[h]
	\caption {
		Calculated bond lengths of A-O (\AA) and B-O (\AA) in the bulk cubic crystal structures of \textit{A}TiO$_3$, (\textit{A}=Ca, Sr, Ba).
	}
	\label{table:bonds}
	\begin{center}
\begin{tabular}{ccc}
	\hline
System	& \textit{A}-O &\textit{B}-O  \\
	\hline
	\hline
	CaTiO$3$&2.75  &1.95  \\

	SrTiO$_3$& 2.78 & 1.97 \\

BaTiO$_3$	& 2.84 &2.01  \\
	\hline
	\hline
\end{tabular}
 \end{center}
\end{table}
The \textit{A}-O bond is  expected to be highly ionic. However, as the electronegativity of \textit{A} increases, we have to appreciate covalent bonding. For the \textit{B}-O bonding, it is expected to be much stronger than the one arising from \textit{A}-O. The bonding is however covalent as well.
Of the three titanates, CaTiO$_3$ has a signature of a large bulk modulus and equally large values of elastic constants. In principle, the elastic constants of CaTiO$_3$ will be higher than those of SrTiO$_3$ and BaTiO$_3$. This is attributed to its short bonds compared to that of SrTiO$_3$ and BaTiO$_3$. 
Regarding hardness, it is not guaranteed that CaTiO$_3$ will be harder than the other two. Before jumping to this conclusion, we advise that one carries out hardness tests to ascertain this. In the mean time, we take this with a pinch of salt since hardness can not be determined a single property like bulk modulus alone. 
\subsection{Band structure of the titanates when in equilibrium}
\label{sec:equilibrum}

\begin{figure}[H]
	\begin{center}
		\includegraphics[width=\columnwidth]{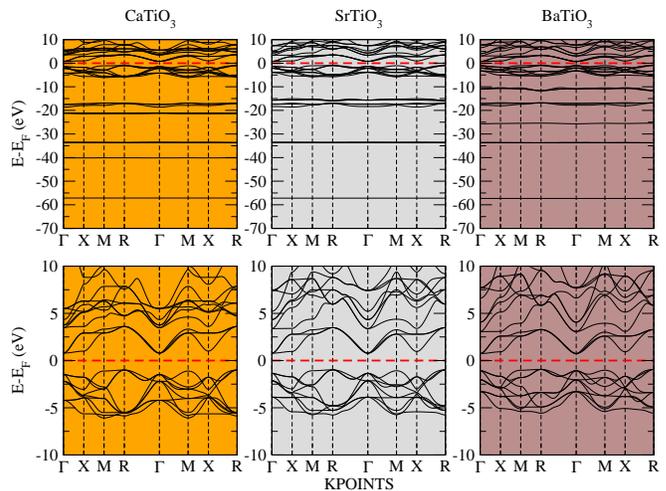}
		\caption{(Color online) Band-structure of the three titanates in the high symmetry cubic
			structure at the equilibrium lattice constant obtained in Table \ref{table:abulkcubic}.
			Top row, the full band structure is shown, including 
			the bands coming from the semi-core states.
			Bottom row displays a zoom highlighting the top of the
			valence bands (O-$2p$ orbitals) and the 
			bottom of the conduction bands (Ti-$3d$ in character). }
		\label{fig:equil}
	\end{center}
\end{figure}
\begin{figure}[H]
	\begin{center}
		\includegraphics[width=\columnwidth]{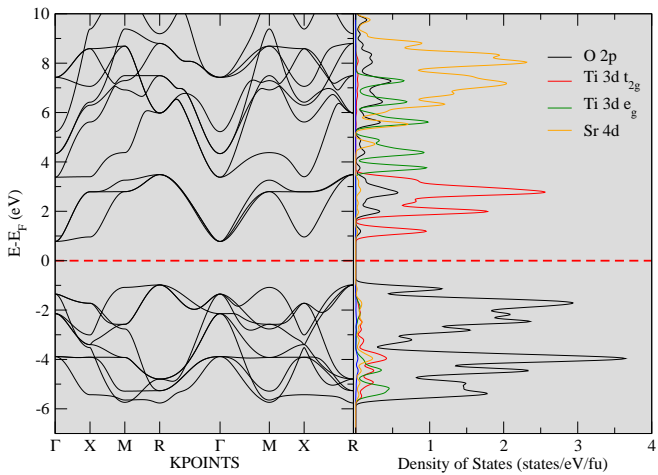}
		\caption{(Color online) Band structure of strontium titanate and its PDOS representation in the cubic
			structure at the equilibrium lattice constant. We use this to illustrate the various orbital composition of the bands drawn. By convention, all the other two remaining titanates in this study have a similar distribution of the orbitals in the band structure. The Fermi level in this case has been set to zero.}
		\label{fig:equill}
	\end{center}
\end{figure}

In Fig.~\ref{fig:equil}, we make a comparison of the band-structures of \textit{A}TiO$_3$, (A=Ca, Sr, Ba) along the selected high symmetry points in the sampled first Brillouin zone for the three cubic bulk perovskite structures. The valence bands are mainly composed of O-2$p$ orbitals that, in the case of the perovskites, is a signature of hybridizations with Ti-3$d$ orbitals. This has been explicitly shown in Fig.~\ref{fig:equill} for the case of SrTiO$_3$, in which all the other two remaining titanates must follow by convention. The top layer indicates that the three materials have an ionic character arising from the well separation of the flat bands approximately having the same energy. The bottom layer however, shows the covalency character due to the dispersive signatures in the bands. The major point here is that, apart from the electrostatic coupling, the ions also communicate due to the overlap of their electron wavefunctions. This trait causes hybridization between the $p$ and $d$ orbitals and the formation of covalent bonds between the \textit{B}-cation and the O-anions.
It is clearly seen that the nine valence bands near the Fermi level are mostly made up of O-$2p$ are triply degenerate. In this case, the splittings observed originate from the crystal field splitting and the electrostatic interaction among the O-$2p$ orbitals.  From Fig.~\ref{fig:equil} and Table \ref{table:gaps}, one can easily tell the variation in the bandgap and bandwidth of the topmost valence region of the three titanates.
The bandgaps and bandwidths of the topmost valence band region are in-principle decreasing with the increased ionic radius in the $\textit{A}$-cation. We found out that the three titanates are insulating in their intrinsic nature, this is attributed to the large ionicity among the three. An indirect bandgap was seen across the three titanates with valence band maximum (VBM) at $R$ while the minima of the conduction  band  was seen at the $\Gamma$ point. In general, all the three titanates follow a given convention in the bandstructure of most oxide perovskites. It is seen also that the bottom of the valence band is made up of  $\sigma$-bonding, while the its top  is made up of the $\pi$-non-bonding traces. The bottom of the conduction band has strong signatures of the $\pi^*$anti-bonding characters as expected~\cite{OK}.

An examination of Table \ref{table:gaps} reveals the significant underestimation of  $E_{gap}$ in this case. This is not a new thing as far as DFT is concerned and it should be taken as such. In this case, our $E_{gap}$ calculations were complicated by the self-interaction error coming from the occupied states in standard DFT~\cite{bandas}, If we consider the "true" band-structure, then, semi-local DFT will have spurious self-interaction only in  occupied states, which in turn over-delocalizes them and thus forces them up in energy and hence the band-gap reduction.

	\begin{table}[H]
	\caption {
		Calculated ($E_{gap}^{theo}$) and experimental ($E_{gap}^{expt}$) bandgaps
		 for CaTiO$_{3}$, SrTiO$_{3}$, and
		BaTiO$_{3}$ in eV for the bulk cubic structure. 
	}
	\label{table:gaps}
	\begin{center}
		
		\begin{tabular}{llc c c}

			\hline
			\hline
			System&$E_{gap}^{theo}$            &
			$E_{gap}^{expt}$ \\
			\hline
			CaTiO$_3$&	1.76                   &
			3.50~\cite{CaTiO3_gap} \\
			
			SrTiO$_3$&	1.74                   &
			3.20~\cite{SrTiO3_gap}\\
			
			BaTiO$_3$&	1.63                   &
			2.83~\cite{Batio_gap}  \\
			\hline
			\hline
		\end{tabular}
	\end{center}
\end{table}

\subsection{Description of the covalent nature in titanates using the PDOS}
\label{sec:orbital}

\begin{figure}[H]
	\begin{center}
		\includegraphics[width=\columnwidth]{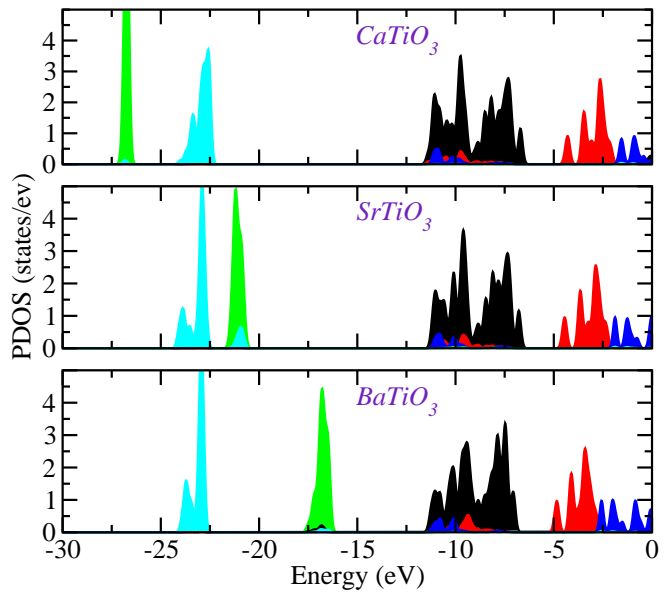}
		\caption{(Color online) Projected density of states of the three perovskites 
			in the cubic structure at the equilibrium lattice constant shown in Table \ref{table:abulkcubic}.
			We show the  projections on the O-$2s$ (cyan), O-$2p$ (black), 
			Ti-$t_{2g}$ (red), Ti-$e_{g}$ (blue), and 
			A-semicore $p$ (green), where A stands for Ca, Sr, or Ba atoms. The conduction band in this case starts from -5eV upwards. }
		\label{fig:eql}
	\end{center}
\end{figure}

%
We performed a further analysis in the covalency nature of the bonds in the three titanates arising from the PDOS. As a simple inspection in Fig.~\ref{fig:eql}, it very easy to see that the composition the valence band is made mostly of O-$2p$ character while the bottom of the conduction band is populated by the Ti-3d orbitals ($t_{2g}$). 
It will be prudent to mention that the electrostatics  on the $Ti$-cation site splits the five-fold degenerate $d$ states into two as shown in Fig.~\ref{fig:eql} ($t_{2g}$ and the $eg$). The
$eg$ group is doubly degenerate and corresponds to the $d$ orbitals whose wave functions have angular symmetry $(x^2- y^2 )/r^2$ and $(3z^2-r^2 )/r^2$ . While the triply degenerate
$t_{2g}$ group corresponds to the $d$ orbitals $(xy/r^2)$, $(xz/r^2)$ and $(yz/r^2)$ ~\cite{PARIDA2018133}. We can clearly see from the PDOS that hybridization occurs between $p$ and $d$ states and thus conforming to covalency. From all the aforementioned similarities, we can see a sharp contrast on the $A$-cation peak with respect to the $O$-states in Fig.~\ref{fig:eql}. As the ionic radius in $A$-cation increases, its peak moves from being  lower energy in CaTiO$_3$ to hybrid in SrTiO$_3$ and finally in higher energies in BaTiO$_3$.
In terms of bonding, ionicity and covalency are depicted in these three titanates. Covalency tends to be more significant if at all we have B-site occupied by a transition element as it is in this case. It is split into both $\sigma$ (A/B-O) and $\pi$ (B-O) bonding as explained in Sec.~\ref{sec:equilibrum}.
\subsection{Interplay of lattice distortion and bandgap in the titanates}
\label{sec:lattice_distortion}
\subsubsection{Application of a uniform tensile strain on the 3 lattices of ATiO$_3$, (A=Ca, Sr, Ba)}
\label{sec:strain}
\onecolumngrid

\begin{center}
	\begin{figure}[H]
		\includegraphics[width=\columnwidth]{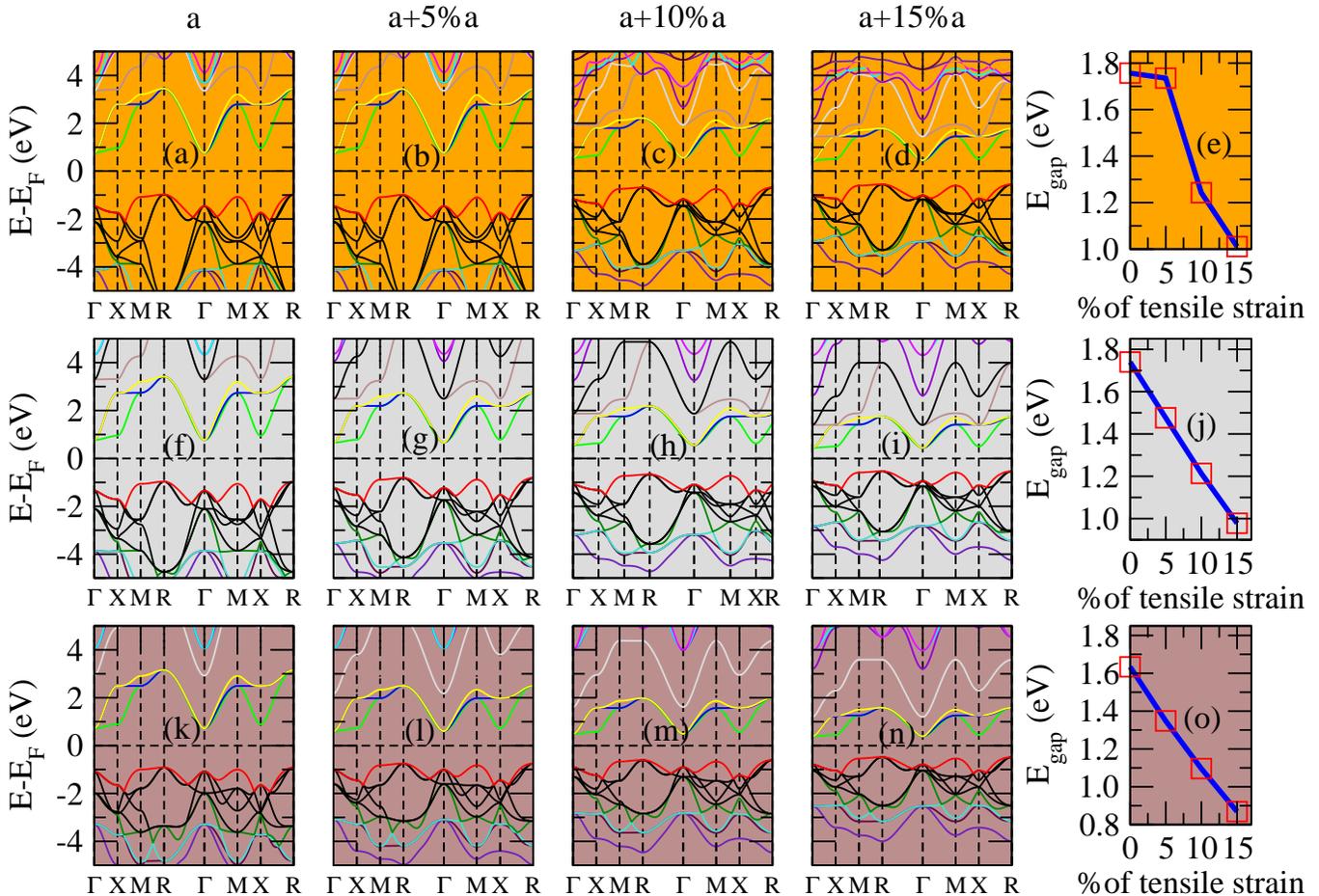}
		\caption{(Color online) Variation of bandgap with tensile strains in the three titanates. Top row (Orange)represents CaTiO$_3$, middle row (grey) represents SrTiO$_3$, while the bottom row (brown) illustrates BaTiO$_3$. $a-d$ illustrates the changes in the bandstructure of CaTiO$_3$ with respective strains. $a$ is the lattice parameter obtained in Table \ref{table:abulkcubic}. $f-i$ show a similar trend as well as $k-n$. $e$, $j$ and $o$ give a schematic variation of the gaps with the values of the strain. It is prudent enough to say that as the lattice is slightly increased, it is possible to tune the bandgap of these titanates to smaller required values.}
		\label{strains1}
	\end{figure}
\end{center}
\twocolumngrid
The main reason why these kind of materials are popular is because their  properties can be altered in a precise manner to obtained any desired feature. In this section, we performed lattice distortions and monitored the bandgaps to ascertain the variation between the bandgap and the lattice parameter in the three titanates.
When a tensile strain was applied, the system had less repulsion and the bands became smaller as seen in Fig.~\ref{strains1}. The gap between A-$d$ orbitals and the $eg$ increased due to the decreased crystal field splitting. This meant that it was very easy to dislodge an electron from the top of the  valence band due to decreased binding force. This explains the decrease in the gap. Degeneracy was observed between $\Gamma$ and $M$ forming a $two$ and a $one$ in the $t_{2g}$.

\subsubsection{Application of a uniform compressive strain on the 3 lattices of \textit{A}TiO$_3$, (A=Ca, Sr, Ba)}
\label{sec:compressive}
\onecolumngrid

\begin{center}
	\begin{figure}[H]
		\includegraphics[width=\columnwidth]{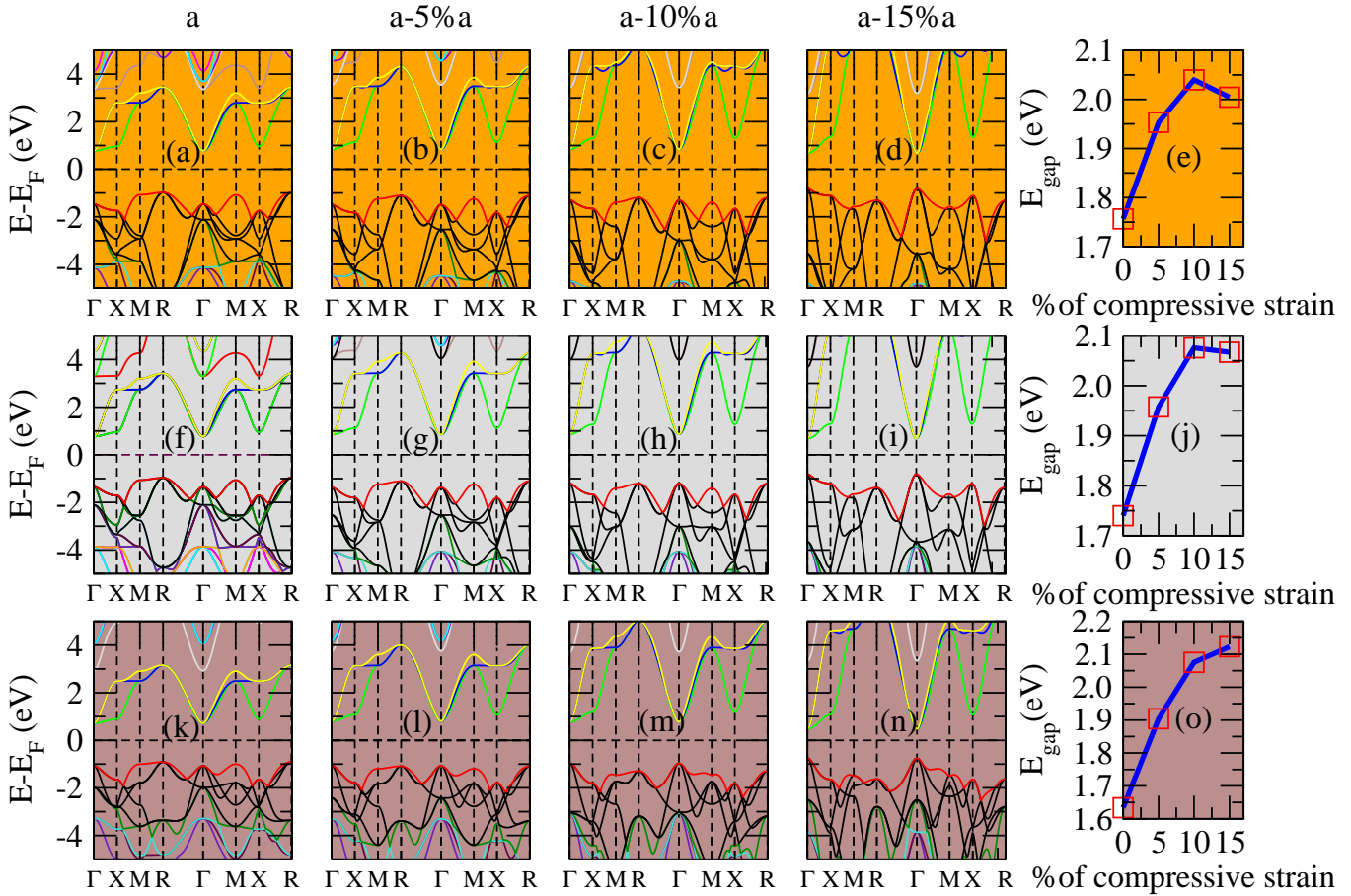}
		\caption{(Color online) Variation of bandgap with respect to compressive strains in the three titanates. Top row (Orange)represents CaTiO$_3$, middle row (grey) represents SrTiO$_3$, while the bottom row (brown) illustrates BaTiO$_3$. $a-d$ illustrates the changes in the bandstructure of CaTiO$_3$ with respective strains. $a$ is the lattice parameter obtained in Table \ref{table:abulkcubic}. $f-i$ show a similar trend as well as $k-n$. $e$, $j$ and $o$ give a schematic variation of the gaps with the values of the strain. It is prudent enough to say that as the lattice is slightly decreased, it is possible to tune the bandgap of these titanates to large required values.}
		\label{comps}
	\end{figure}
\end{center}
\twocolumngrid
It was noted that when the system underwent a compression, the bands enlarged as shown in Fig.~\ref{comps}. At the $\Gamma$ point for instance, the distance between the $t_{2g}$ and $eg$ is enormous despite both being degenerate between $R$ and $\Gamma$. This is attributed to the strong coulombic repulsion. Degeneracy was broken at points $X$  where two independent bands were observed in the $t_{2g}$. The $eg$ was seen to get attracted to the \textit{A}-$d$ (\textit{A}=Ca, Sr, Ba). One more thing that came out clearly was the increased band gap. As the lattice decreased, the interatomic distance decreased and thus increasing the binding force between the valence electrons and the parent atoms. In such a case, the systems needed more energy to make the electrons free and this accounts for the large gap observed. Using a compression beyond 10\% set precedent of the novel  cooperative Jahn-Teller effects and this explains the behaviour of Figs.~\ref{comps}e,~\ref{comps}j and ~\ref{comps}o.

\section{Conclusion and future works}
\label{sec:conclusions}
First and foremost, it is worth noting the remarkable power of first principles calculations since our results have been seen to be in full agreement with existing experimental values reported by authors cited herein. We appreciate the fact that we now understand the nature of bonding in the selected \textit{A}TiO$_3$ crystalline structures to be an admixture of  covelent and ionic character. Apart from the lattice differences in the \textit{A}TiO$_3$  titanates, we see from the PDOS that the contribution of the $A$ cation is correlated to the atomic number of a given titanate. Also discussed was the various orbital contributions in the bandstruture with O-2$p$ and Ti-3$d$ $t_{2gs}$ being dominant near the Fermi in all the three materials. We have equally seen that the bandstructure is highly sensitive of the lattice distortion as predicted by Dalven~\cite{delvin} many years ago in semiconductors. In our case, we found a lattice bandgap relationship of the form $E_{gap}\propto \frac{1}{a^x}$
with $2.19<x<3.1$.  In the case of compressive strains, we predict that hopefully similar effects may be achieved by sizable pressure applications on the titanates.
Although we have mentioned that our results are comparable, in future, it may be prudent to use an improved exchange-correlation energy functional like a hybrid. Equally, phonon dispersions for these titanates under strains should be computed to analyse the possibility of instabilities. 
The study of cooperative Jahn-Teller distortions have not been studied herein and this should open an avenue for new research in this line.  
\section*{ACKNOWLEDGMENT} 
\label{sec:Acknowledgement} 
We acknowledge George Manyali for the extremely useful discussions we had regarding the ATiO$_3$ class of materials. The authors also gratefully acknowledge the computer resources, technical expertise, and assistance provided by the Centre for High Performance Computing (CHPC-MATS862 \& MATS0712), Cape Town, South Africa. 
\bibliography{james}
\end{document}